\documentclass{osa-article}
\usepackage{leitian_v2}
\usepackage{upgreek}

\usepackage{amsmath,amssymb,amsfonts}
\usepackage{graphicx}
\usepackage{textcomp}
\usepackage{comment}
\usepackage[ruled, lined, commentsnumbered, longend]{algorithm2e}
\usepackage{hyperref}

\journal{oe}

\begin{document}

\title{Large-scale holographic particle 3D imaging with the beam propagation model}

\author{Hao Wang,\authormark{1} Waleed Tahir \authormark{1}, Jiabei Zhu,\authormark{1}, and Lei Tian\authormark{1,*}}

\address{\authormark{1}Department of Electrical and Computer Engineering, Boston University, Boston, MA 02215, USA}

\email{\authormark{*}leitian@bu.edu} 



\begin{abstract*}
We develop a novel algorithm for large-scale holographic reconstruction of 3D particle fields. 
Our method is based on a multiple-scattering beam propagation method (BPM) combined with sparse regularization that enables recovering dense 3D particles of high refractive index contrast from a single hologram. 
We show that the BPM-computed hologram generates intensity statistics closely matching with the experimental measurements and provides up to 9$\times$ higher accuracy than the  single-scattering model.
To solve the inverse problem, we devise a computationally efficient algorithm, which reduces the computation time by two orders of magnitude as compared to the state-of-the-art multiple-scattering based technique. 
We demonstrate the superior reconstruction accuracy in both simulations and experiments under different scattering strengths.
We show that the BPM reconstruction significantly outperforms the single-scattering method in particular for deep imaging depths and high particle densities.
\end{abstract*}

\section{Introduction}
\label{sec:introduction}
Single-shot holographic particle 3D imaging is fundamental to many important applications, such as flow cytometry~\cite{cheong2009flow}, biological sample characterization~\cite{moon2009automated, su2012high}, and flow measurement~\cite{tian2010quantitative}. 
The technique works by first recording a 2D intensity measurement from a particle volume under coherent illumination and then estimating the 3D refractive index distribution. 
The problem is challenging because of the dimensional mismatch between the single 2D measurement and the unknown 3D object and the ``phaseless'' measurement, both of which make the inverse problem severely ill-posed, especially for large-scale problems. 
Furthermore, multiple scattering effects become significant as the particle density and the refractive index contrast increase, which necessitates a nonlinear forward model to accurately describe the image formation process.

Traditional holographic particle 3D reconstruction is based on the linear single-scattering model derived from the first Born approximation method (FBM)~\cite{born1999optics}.
The most widely used algorithm is known as the ``back-propagation method''~\cite{tian2010quantitative}, which is equivalent to apply the pseudo-inverse of the (3D-to-2D) single-scattering forward operator to the captured hologram. 
To improve the reconstruction accuracy, compressive holography~\cite{brady2009compressive} has been developed that supplements the FBM forward model with a sparsity constraint and solves the 3D reconstruction problem by an iterative optimization procedure. 
This approach is particularly effective in alleviating the twin-image artifacts arising~\cite{brady2009compressive} and improving the robustness to unaccounted multiple-scattering effects at high scattering densities~\cite{chen2015empirical}.
However, these methods are fundamentally limited by the underlying single-scattering assumption, which is  valid only for weakly scattering objects.
The model gradually breaks down as the particle density and/or the refractive index contrast increase.
The multiple-scattering effects result in a model mismatch, which in turn limits the reconstruction accuracy by these techniques.

Several multiple-scattering models have been developed, such as iterative Born series~\cite{kamilov2016recursive,tahir2019holographic}, contrast source inversion~\cite{van1997contrast}, discrete-dipole approximation~\cite{draine1994discrete,unger2019versatile,mudry2012electromagnetic,zhang2016far}, and series expansion with accelerated gradient descent on Lippmann–Schwinger equation~\cite{liu2017seagle}.   
However, computational challenges typically restrict these methods to be used only for small-scale problems.
This is because they involve iteratively estimating the internally scattered fields within the object volume for modeling the multiple-scattering process, which incurs a computation cost that  grows rapidly as the object size, and thus is not ideal for our intended applications containing on the order of 100 million voxels in the object volume.
Recently, the multi-slice beam propagation method (BPM)~\cite{ kamilov2016optical,tian20153d, chowdhury2019high,chen2020multi} has emerged as an attractive, computationally efficient multiple-scattering model.
The utility of the BPM has been demonstrated for reconstructing 3D refractive index distributions using tomographic measurements from multiple interferometric complex-field measurements~\cite{ kamilov2016optical} or intensity-only measurements~\cite{tian20153d, chowdhury2019high,chen2020multi}.
Our goal here is to critically examine the utility of the BPM to handle the inversion from a single in-line hologram, which inherently suffers from more severe missing-cone artifacts~\cite{tam1981tomographical,lim2015comparative} and greater dimensional mismatch as compared to the previous tomography studies. 

We demonstrate a novel 3D  reconstruction algorithm that combines a BPM multiple-scattering forward model and a sparsity prior.  
The accuracy of our forward model is quantified by comparing the simulation with the experiments.
We show that the BPM-computed intensity statistics closely match with the experimental data at different scattering densities, and provide up to 9$\times$ higher accuracy in predicting the hologram contrast than that from the FBM.
Benefited from the high computational efficiency, the BPM reduces the computational complexity by $N_z$ times (where $N_z$ is the number of axial slices of the object volume), as compared to the state-of-the-art multiple-scattering model-based holographic particle reconstruction method~\cite{tahir2019holographic}. 
We demonstrate our proposed method's superior 3D imaging performance in both simulation and experiments under different scattering densities and refractive index contrast. 
To quantify the improvement by the multiple-scattering model over the single-scattering model, we compare our method with the compressive holography method. 
We show that the BPM reconstruction significantly outperforms the FBM in particular at deep imaging depths and for high particle densities.

\section{Method}
\label{method}
Our reconstruction algorithm solves an $\ell_1$-regularized least-squares problem, in which the data fidelity term measures the difference between the BPM-estimated and captured holograms. 
Below, we describe our forward model and the reconstruction algorithm.

\subsection{Forward model}
\label{forward}
\begin{figure}[ht!]
\centerline{
 \includegraphics[width=1\textwidth]{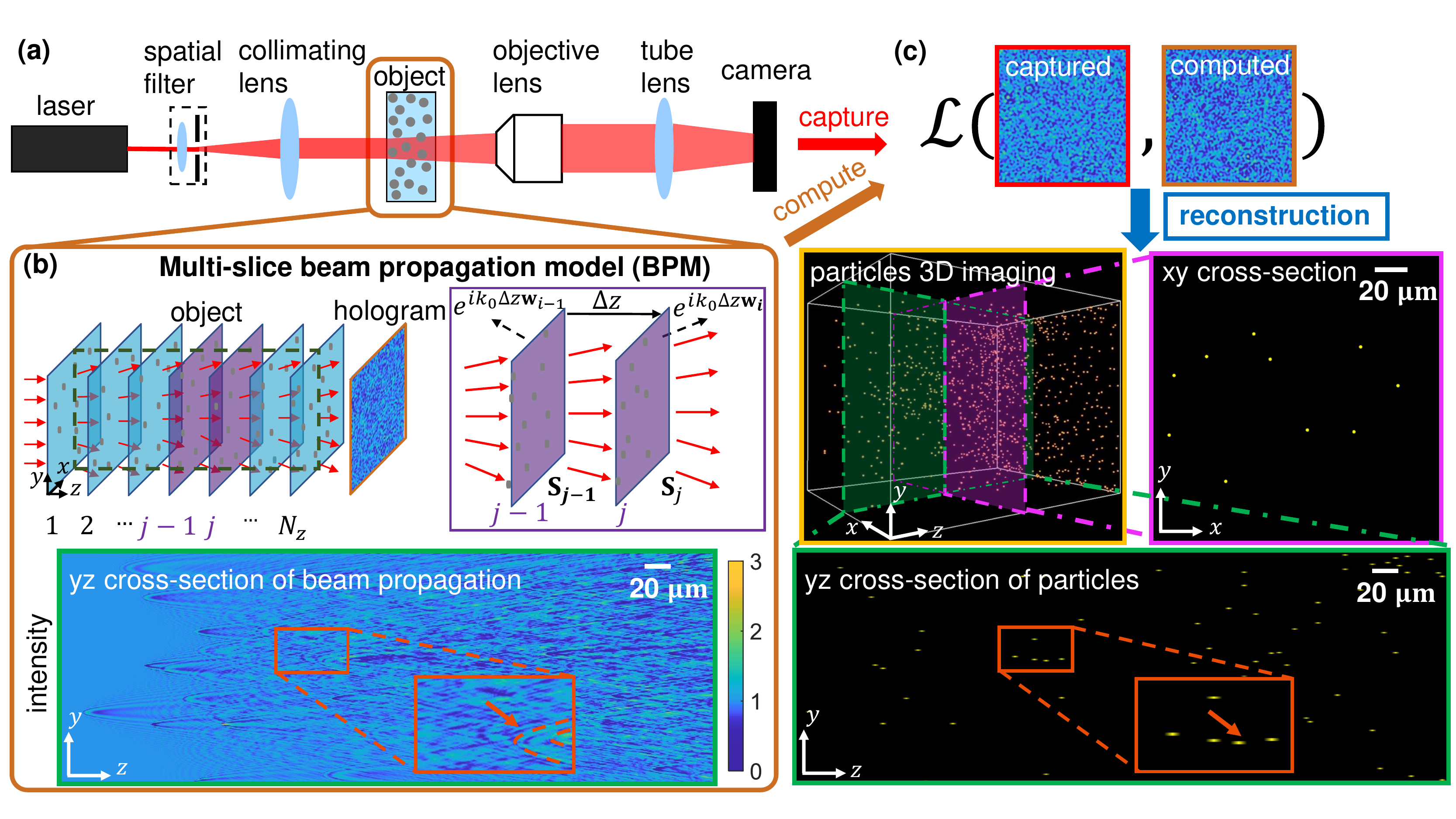}}
 \caption{Particle 3D imaging from a single in-line hologram using BPM. (a) Experimental setup. 
 (b) Top: BPM involves successive propagation and refraction operations to transform the scattered field from one object slice to the next. The hologram is computed as the squared magnitude  of the scattered field. Bottom: the $yz$ cross-section of the BPM computed intensity distribution of light passing through a particle volume. The zoom-in region illustrates the BPM-predicted inter-particle multiple scattering.
 (c) Top: 3D reconstruction by solving a sparsity-driven minimization problem. Bottom: the reconstruction from the hologram in (b).}
 \label{fig:1}
\end{figure}

Our imaging geometry is shown in Fig.~\ref{fig:1}(a). 
A plane wave passes through a particle volume and the interference pattern resulting from the scattered fields are captured as the hologram. 
This multiple-scattering image-formation process is modeled as, 
\be
\hat{\bI}=|\bS(\bw)|^2,
\label{equ1}
\ee
where $\hat{\bI}\in\mathbb{R}^M$ denotes the vectorized, BPM computed intensity hologram containing $M$ pixels. $\bw\in\mathbb{R}^N = \bn-n_0$ represents the vectorized 3D refractive index contrast distribution that contains $N$ voxels, and is defined by the difference between the particle's index distribution $\bn$ and the constant background index $n_0$. The particle's refractive index is assumed to be real-valued since the absorption is negligible.
$\bS: \mathbb{R}^N\rightarrow\mathbb{C}^M$ is the BPM operator that maps the 3D refractive index to the 2D complex field at the sensor plane. 

The BPM is computed by a recursive sequence, as illustrated in Fig.~\ref{fig:1}(b). 
It approximates the 3D object as $N_z$ infinitesimally thin axial planar slices along the optical axis, each modeled as a 2D phase mask separated equally by a uniform medium.
The field is then computed by a sequence of diffraction and refraction operations. 
Specifically, the field exiting the $j\mr{th}$ slice $\bS_j(\bw)$ is related to the field from the $(j-1)\mr{th}$ slice $\bS_{j-1}(\bw)$ by
\be
\bS_j(\bw)= \mr{diag}(\bp_j(\bw_j))\bH\bS_{j-1}(\bw).
\label{equ2}
\ee
where the $\bS_j: \mathbb{R}^M\rightarrow\mathbb{C}^M$ is the local BPM operator that maps the $j\mr{th}$ phase screen $\bw_j$ to the 2D complex field after the $j$th slice.

The implementation of the BPM requires an initial condition, for which we set the initial field as a constant-valued on-axis plane wave $\bS_0(\bw) = \mathbf{1}$.

The diffraction operator for a propagation distance $\Delta z$ is denoted by $\bH$ and computed by
$\bH = \bF^H\mr{diag}(\bh)\bF$, where $\bF$ and $\bF^H$ are the 2D discrete Fourier and inverse Fourier transform matrices respectively, $\large{\cdot}^H$ denotes matrix Hermitian transpose, $\mr{diag}(\bh)$ is a diagonal matrix formed by the vector $\bh$ as the main diagonal and is implemented by element-wise multiplication, and $\bh$ represents the vectorized 2D angular spectrum  transfer function $h(p,q)$:
\be
h(p,q)=\exp\{i\Delta z\sqrt{k^2-(p\Delta k)^2-(q\Delta k)^2}\}\mathcal{P}(p,q),
\label{equ3}
\ee
where $p,q$ index the 2D wave vector in the $k$-space,
$k=k_0n_0$ is the wave-number in the background medium, $k_0={2\pi}/{\lambda_0}$, $\lambda_0$ is the wavelength in vacuum, and $\Delta k$ is the sampling interval in the $k$-space.
$\mathcal{P}$ represents the low-pass filter due to the evanescent cutoff.

The refraction operator is implemented by element-wise multiplications between the diffracted field and the accumulated phase $\bp_j(\bw_j)=e^{jk_0\Delta z\bw_j}$, where $\bw_j$ denotes the discretized 2D refractive index map in the $j\mr{th}$ slice. 

The hologram is the intensity of the exit field from the last slice at $N_z$ low-pass filtered by the pupil function of the imaging system.

The computation of the BPM thus requires implementing Eq.~\eqref{equ2} $N_z$ times. 
The computational complexity of the BPM is $O(N\mr{log}(N_x N_y))$, as set approximately by $2N_z$ times evaluations of 2D FFT, where the total number of object voxels is $N=N_xN_yN_z$, and $N_x$ and $N_y$ are the number of pixels in the $x$ and $y$ directions, respectively.
For comparison, the computational complexity of the Born series model is $O(N_zN\mr{log}(N_x N_y))$~\cite{tahir2019holographic}.
Thus, the BPM reduces the computational complexity by $N_z$ times, which becomes significant when the object depth is large.  

To illustrate the multiple scattering effects modeled by the BPM, we show an example $yz$ cross-section of the intensity distribution computed from a particle volume in Fig.~\ref{fig:1}(b). 
In the zoom-in region, complex interference patterns are visible due to multiple particles in the close proximity, which introduces strong multiple scattering effects~\cite{lim2018learning}.

\subsection{Inverse problem}
\label{inverse}
\subsubsection{Problem formulation}
We formulate the 3D reconstruction as a minimization problem:
\be
\hat{\bw} =\underset{\bw\in\mathbb{R}^N} {\operatorname{argmin}}\{\mathcal{D}(\bw)+\tau \mathcal{R}(\bw)\},
\label{equ5}
\ee
where $\mathcal{D}$ is the data fidelity term and $\mathcal{R}$ is the regularization term.
The parameter $\tau>0$ controls the amount of regularization.
The data fidelity term is given by
\be
\mathcal{D}(\bw)\triangleq\frac{1}{2}\|\bI-\hat{\bI}\|_2^2,
\label{equ6}
\ee
where $\bI$ is the captured hologram and $\|\cdot\|_2$ is the $\ell_2$ norm. 

The regularization term is the $\ell_1$ norm of the refractive index distribution
\be
\mathcal{R}(\bw)\triangleq\|\bw\|_1,
\label{equ7}
\ee
which promotes the sparsity of the reconstructed object. 

The minimization Eq.~\eqref{equ5} is not a trivial task. 
The primary difficulty stems from that the data fidelity term $\mathcal{D}$ involves a nonlinear forward operator $\bS$ and the  regularization term $\mathcal{R}$ is non-smooth. 
We next present a novel algorithm based on the proximal-gradient descend technique, which extends \cite{kamilov2016optical} to intensity-only measurement.

\subsubsection{Computation of the gradient}
The crucial step is the gradient computation for $\mathcal{D}$ with respect to $\bw$, as summarized in Algorithm~\ref{alg1} and briefly explained below. Additional details are provided in the Supplementary material. 

\begin{algorithm}
\caption{Gradient computation: $\nabla[\mathcal{D}(\hat{\bw})]^H$}
\SetKwFunction{isOddNumber}{isOddNumber}
\label{alg1}
\SetKw{input}{input:}
\SetKw{output}{output:}
\SetKw{Algorithm}{Algorithm:}
\KwIn{incident field $\bS_0(\bw)$, measured hologram $\bI$, and initial estimate of refractive index distribution $\hat{\bw}^0$.}
\KwOut{the gradient $[\nabla \mathcal{D}(\hat{\bw})]^H$.}
\Algorithm{}\\
\Indp 1) Compute the exit field $\bS_{N_z}(\hat{\bw})$ using the BPM recursion in Eq.~\eqref{equ2}; estimate the hologram $\hat{\bI}$ by Eq.~\eqref{equ1}; keep all the intermediate fields $\bS_j(\hat{\bw})$ in memory.\\
2) Compute $\br_{N_z} = \mr{diag}(\bS_{N_z}(\bw))(\hat{\bI}-\bI)$ and set $s_{N_z}=0$.\\
3) Compute $s_0=[\frac{\partial }{\partial \bw}\bS_{N_z}(\hat{\bw})\big]^H \br_{N_z}$ using the following iterative procedure for slices $j=N_z,...,1$\\
     \Indp 3a) $s_{j-1} = \underbrace{s_j + \big[\frac{\partial}{\partial \bw}\bp_{j}(\hat{\bw}_j)\big]^H\mr{diag}(\overline{\bH\bS_{j-1}(\hat{\bw}})\br_{j}}_\textrm{update intermediate field gradient slice by slice},$\\
     3b) $\br_{j-1} =\underbrace{ \bH^H\mr{diag}(\overline{\bp_{j}(\hat{\bw}_j)})\br_{j}}_\textrm{backpropagate the field residual}$.\\
\Indm 4) Return: $\nabla [\mathcal{D}(\hat{\bw})]^H = 2\mr{Re}\{s_0\}$.
\end{algorithm}

First, we take the derivative of $\mathcal{D}(\bw)$ with respect to $\bw$ and rearrange it into a column vector 
\be
\begin{split}
\big[\frac{\partial{\mathcal{D}(\bw)}}{\partial \bw} \big]^H
&=\big[\frac{\partial{\hat{\bI}}}{\partial \bw}\big]^H\br,
\end{split}
\label{equ10}
\ee
where $\br\triangleq\hat{\bI}-\bI$ is the residual between the estimated and captured holograms.
The estimated measurement can be written as $\hat{\bI} = |\bS_{N_z}(\bw)|^2
    =\mr{diag}(\bS_{N_z}(\bw))\overline{\bS_{N_z}(\bw)},$
where {\Large$\overline{\cdot}$} denotes taking the complex conjugate.
Thus, the gradient of the data fidelity term is
\be
[\nabla \mathcal{D}(\bw)]^H = 2\mr{Re}\{\big[\frac{\partial \bS_{N_z}(\bw)}{\partial \bw}\big]^H \mr{diag}(\bS_{N_z}(\bw))\br\}.
\label{equ15}
\ee

To derive a tractable algorithm for computing Eq.~\eqref{equ15}, we apply the recursive BPM formula in Eq.~\eqref{equ2} and compute the local gradient at the $j$th slice: 
\be
\begin{split}
\big[\frac{\partial }{\partial \bw}\bS_{j}(\bw)\big]^H
&=\big[\frac{\partial}{\partial\bw}\big(\mr{diag}(\bp_j(\bw_j))\bH\bS_{j-1}(\bw) \big)\big]^H\\
& = \big(\frac{\partial}{\partial \bw}\bp_j(\bw_j)\big)^H\mr{diag}(\overline{\bH\bS_{j-1}(\bw)}) + \big(\frac{\partial}{\partial \bw}\bS_{j-1}(\bw)\big)^H \bH^H \mr{diag}(\overline{\bp_j(\bw_j)}).\\
\end{split}
\label{equ16}
\ee

Since the input plane wave  does not depend on $\bw$, so for the initial condition $j=0$, we have
\be
\big[\frac{\partial}{\partial\bw}\bS_0(\bw)\big]^H=0.
\label{equ18}
\ee
Based on Eq.~\eqref{equ16} and the initial condition in Eq.~\eqref{equ18}, we obtain a practical implementation of Eq.~\eqref{equ15} to calculate the gradient of the data fidelity term, as summarized in Algorithm~\ref{alg1}. 

Intuitively, the gradient computation is similar to the ``error backpropagation'' algorithm used in deep neural networks~\cite{kamilov2016optical}. 
The algorithm iterates between two major steps, including update the intermediate field gradient slice-by-slice (first term in Eq.~\eqref{equ16}) and backpropagate the slice-wise field residual (second term in Eq.~\eqref{equ16}).
Since this gradient computation takes the same recursive procedure as the forward BPM model, its computation complexity is also $O(N\mr{log}(N_x N_y))$. 

\subsubsection{Reconstruction algorithm}
Our algorithm is summarized in Algorithm~\ref{alg2}, that reconstructs the refractive index $\bw$ from a single hologram based on the proximal gradient algorithm. Conceptually, this algorithm is similar to the fast iterative shrinkage/thresholding algorithm (FISTA)~\cite{beck2009fast}, which is widely used to minimize objective function that consist of the sums between a smooth and a non-smooth term.
\begin{algorithm}
\caption{Minimizes: $\mathcal{D}(\bw)+\tau\mathcal{R}(\bw)$}
\label{alg2}
\SetKwFunction{isOddNumber}{isOddNumber}
\SetKw{input}{input:}
\SetKw{set}{Set:}
\SetKw{repeat}{Repeat}
\SetKw{until}{Until}
\SetKw{return}{Return}

\KwIn{measured hologram $\bI$, initial guess $\hat{\bw}^0$, step size $\gamma$, and regularization parameter $\tau$.}
\set{ $t \gets 1, \bs^0 \gets \hat{\bw}^0, q_0 \gets 1$}\\
\repeat{}\\
\Indp 
$\bba^t \gets \bs^{t-1}- \gamma \nabla \mathcal{D}(\bs^{t-1})$ \\
$\hat{\bw}^t \gets \mr{prox}_{\gamma\tau\|\cdot\|_1}(\bba^t)$\\
$q_t \gets \frac{1}{2}\big(1+\sqrt{1+4q_{t-1}^2}\big)$\\
$\bs^t \gets \hat{\bw}^t + ((q_{t-1}-1)/q_t)(\hat{\bw}^t-\hat{\bw}^{t-1})$\\
$t \gets t+1$\\
\Indm \until{} stopping criterion\\
\return{} estimate of the refractive index $\hat{\bw}^t$
\end{algorithm}

A major component of this algorithm is the proximal operator for the $\ell_1$-regularizer
\be
\mr{prox}_{\gamma\tau\|\cdot\|_1}(\bba)\triangleq\underset{\bw\in\mathbb{R}^N}{\operatorname{argmin}}\{\frac{1}{2}\|\bw-\bba\|^2_2+\gamma\tau \|\bw\|_1\},
\label{equ19}
\ee
where $\bba$ and $\gamma$ are explained in Algorithm~\ref{alg2}. 
This proximal operator has a closed-form solution, known as soft-thresholding, see in Supplementary material.

The parameters in Algorithm~\ref{alg2} are set as follows. We set the initial guess $\hat{\bw}^0 = \mathbf{0}$, and the step size $\gamma=5\times10^{-6}$. 
The stopping criterion is the maximum iteration number to be 200-300. 
The regularization parameter $\tau$ is tuned under different scattering conditions. 
When scattering is stronger, $\tau$ is set larger.
The reconstruction is found to be insensitive to the fine-tuning of $\tau$.

\section{Evaluation of BPM forward model accuracy in large scale}
\label{sec:3}
\subsection{Intensity statistics analysis}
We first validate the forward model accuracy by comparing the BPM-computed holograms with experimental measurements. 
In practice, we assess the accuracy based on analyzing the intensity statistics under different scattering conditions, since the ground-truth particle positions in the experiments are not known.

\begin{figure}[ht!]
 \centering 
 \includegraphics[width=0.85\textwidth]{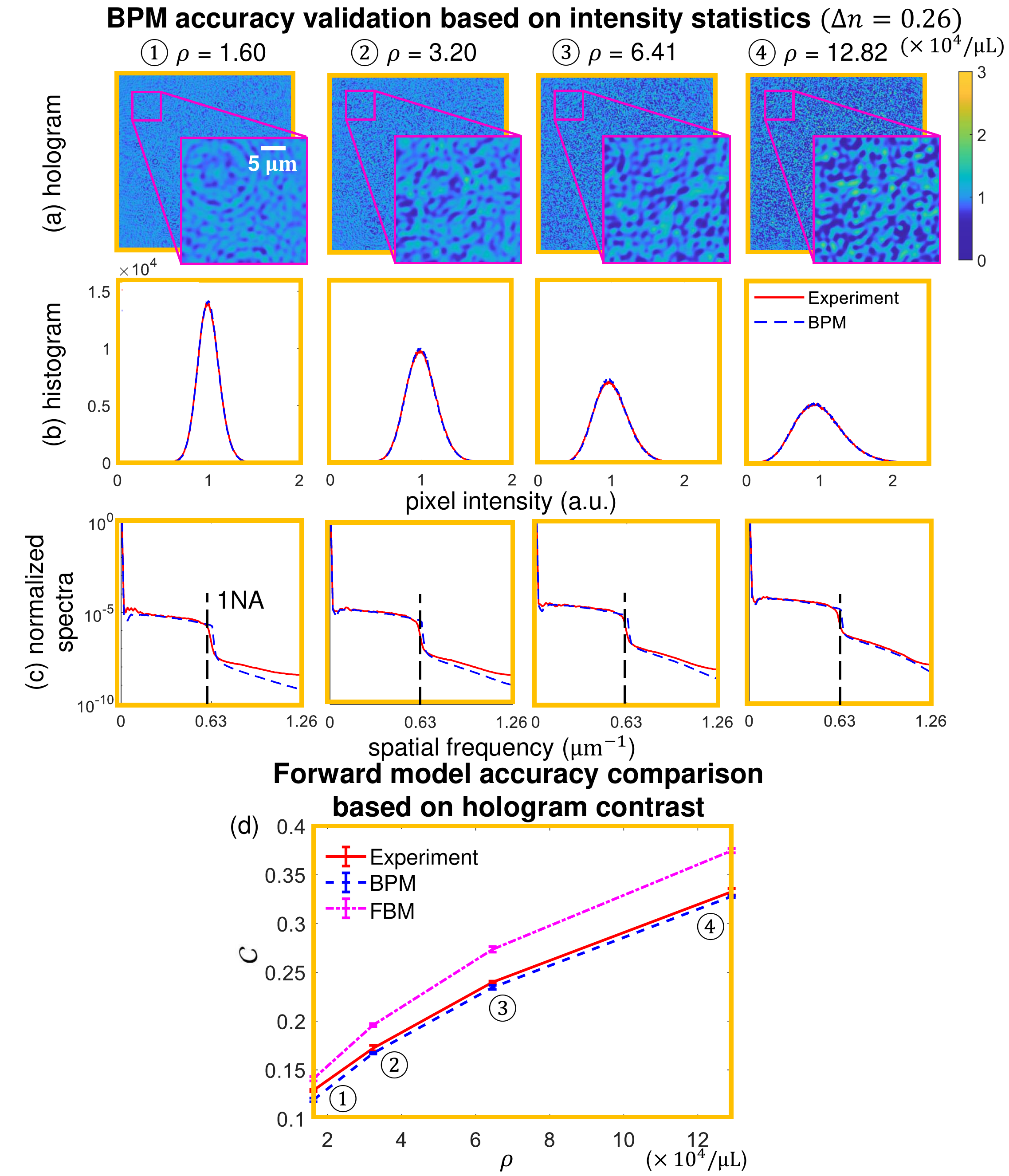}
 \caption{\label{fig:2} BPM model validation based on intensity statistics.
 (a) Example holograms computed from the BPM at different particle densities with a fixed refractive index contrast $\Delta n =0.26$. 
 (b) The intensity histograms and (c) normalized intensity spectra of BPM-computed (in dashed blue) and experimentally captured (in solid red) holograms. The results are averaged azimuthally. (d) Comparison of BPM and FBM accuracy based on hologram contrast.}
\end{figure}

We perform simulations using parameters that match with the experiments. 
More details about the experiments are provided in Supplementary material. 
Holograms are computed at four particle densities $\rho$, including $1.60, 3.20, 6.41, 12.82\times10^{4} /\upmu \mr{L}$, corresponding to $250, 500, 1000, 2000$ particles in a $176.64 \times 176.64 \times 500~\upmu \mr{m}^3$ volume. 
$1~\upmu \mr{m}$ particles are randomly placed in 3D positions. 
The refractive index of the particle $n$ and the background medium (water) $n_0$ is 1.59 and 1.33, respectively, and the contrast is $\Delta n=0.26$. 
To simulate 3D refractive index contrast distribution $\bw$, the voxels inside the particles are assigned with the constant $\Delta n$, and the rest of the background voxels are assigned with zero.
The lateral sampling size $\Delta x, \Delta y$ are both $0.1725\upmu\mr{m}$, and the axial sampling size $\Delta z$ is $\lambda/16=0.0297\upmu\mr{m}$, where $\lambda=\lambda_0/n_0$ and $\lambda_0=0.632\upmu\mr{m}$. 
The resulting object size in our forward model is $1024\times1024\times16840$ voxels.
Example computed holograms at four particle densities are shown in Fig.~\ref{fig:2}(a).
As expected, characteristic fringe patterns from individual particles are still visible at the lowest density case. 
As the density increases, the holograms gradually become partially developed speckle patterns. 

First, we analyze the BPM's accuracy by comparing the histograms of the computed hologram with the captured hologram at each particle density. 
As shown in Fig.~\ref{fig:2}(b), the histograms match well at all four densities. 

Next, we assess the BPM's accuracy in the spatial frequency domain. 
We calculate the normalized spectra of the computed and captured holograms. 
As shown in Fig.~\ref{fig:2}(c), the frequency components of the computed holograms closely match with the experimental measurements within the NA of the imaging system.

Finally, we compare the accuracy of the BPM and FBM based on the hologram contrast $\mathcal{C} = \frac{\sigma}{\mu}$, where $\sigma$ and $\mu$ are the standard deviation and mean of the hologram, respectively. 
Briefly, the FBM is a linear model that assumes a weakly scattering object, and the resulting scattered field is linearly related to the scattering potential $V(x,y,z) = \frac{1}{4\pi}k_0^2[n(x,y,z)^2-n_0^2]$. Given the plane-wave incident field $U_{in}(x,y,z) = e^{in_0k_0z}$, the total field $U_{\text{FBM}}$ can be approximated as
\begin{multline}
U_{\text{FBM}}(x,y,z)\approx U_{in}(x,y,z) +\iiint G(x-x',y-y',z-z')U_{in}(x',y',z')V(x',y',z') \,dx'\,dy'\,dz'
\label{equ20}
\end{multline}
where the Green's function $G(r) = {\exp(ik_0n_0r)}/{r}$, $n(x,y,z)$ is the refractive index at location $(x,y,z)$ and $r=\sqrt{x^2+y^2+z^2}$~\cite{born1999optics}.
The FBM-estimated hologram is $I_{\text{FBM}} = |U_{\text{FBM}}|^2$.
To perform the computation, the 3D volume is first discretized with voxels in size $\Delta x\times \Delta y \times \Delta z$. 
In our simulation, $\Delta x = \Delta y = 0.1725 \upmu\mr{m}$ and $\Delta z = \lambda/16$. 
The scattering potential is calculated as $\frac{1}{4\pi}k_0^2 \Delta x\Delta y\Delta z(n^2-n_0^2)$ for voxels within the particle, and is zero for the rest of the voxels. 
To compute the scattered field, we treat the discretized volume as a series of 2D slabs. 
The scattered field from a given slab can be efficiently calculated by first multiplying the incident field at the given depth with the scattering potential of the slab and then convolve with the Green's function implemented with the fast Fourier transform (FFT)-based algorithm. 
The total field $U_{\text{FBM}}$ is obtained by summing the incident field at the hologram plane and the total scattered field from all the slabs.

As shown in Fig.~\ref{fig:2}(d), the contrast from the BPM-computed holograms agree well with the experiments. 
The BPM slightly under-estimates the contrast. 
A possible reason is that the BPM approximates the multiple forward scattering by computing the complex field slice-by-slice but ignores backscattering. 
As a result, the BPM may slightly under-estimates the high frequency information, which reduces the contrast in the calculated holograms. As a comparison, the contrast from the FBM-computed holograms are consistently higher than the experimental data.
The contrast discrepancy increases as the particle density.
At the highest density, the discrepancy for the BPM is 0.0048, whereas the FBM is 0.0422, representing a $9\times$ improvement by the BPM. 

Overall, these studies show that the BPM can accurately model multiple scattering in a dense 3D particle field and significantly outperforms the single-scattering model.

\subsection{Sampling distance $\Delta z$ and scattering strength effect in BPM forward model}
\label{larger dz}
The BPM accuracy is primarily influenced by the axial sampling distance $\Delta z$ and the scattering strength of the 3D object.
In the following, we quantitatively evaluate the model accuracy under different axial sampling and scattering conditions (by changing the particle density $\rho$ and refractive index contrast $\Delta n$), while fixing the lateral sampling $\Delta x = \Delta y = 0.1725\upmu\mr{m}$.

\begin{figure}[ht!]
\centering 
\includegraphics[width=1\textwidth]{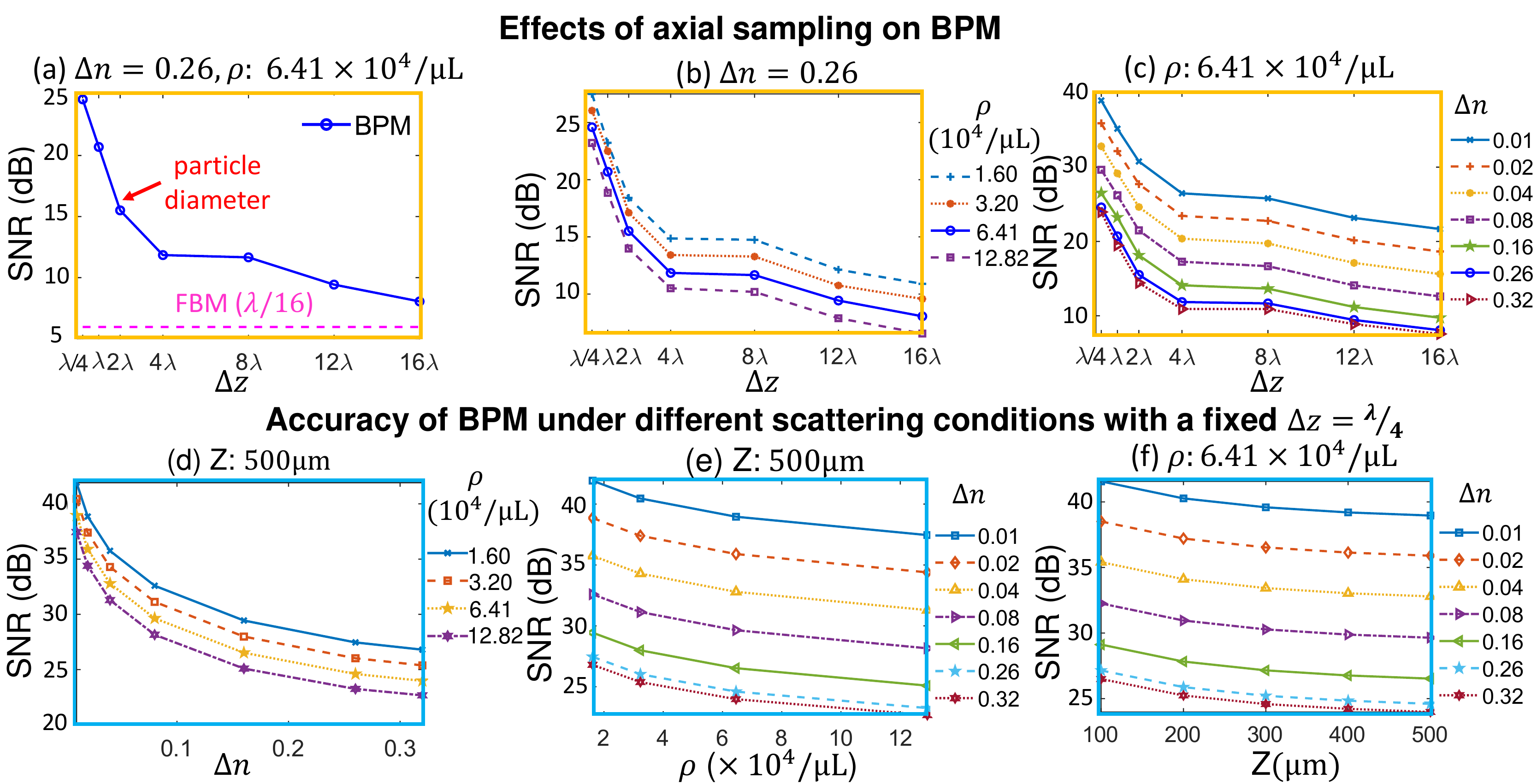}
\caption{\label{fig:3} The effects of axial sampling and scattering strength on BPM. (a) The accuracy of BPM decreases as $\Delta z$ increases. Overall, the BPM significantly outperforms the FBM. 
(b-c) The SNR of the BPM-computed hologram reduces as $\Delta z$ increases under (b) different particle densities and (c) refractive index contrasts.
(d-f) The accuracy of the BPM under different scattering conditions with a fixed $\Delta z=\lambda/4$. 
The SNR reduces when increasing (d) the refractive index contrast, (e) particle density, and (f) total object thickness.}
\end{figure}

In Fig.~\ref{fig:3}(a), the accuracy of BPM is plotted for different $\Delta z$ under the same scattering condition. 
We compare the holograms computed with different $\Delta z$ with the reference $\hat{\bI}_0$ using $\Delta z=\lambda/16$.
The difference is quantified by the signal-to-noise ratio (SNR), $\mr{SNR}=10\log_{10}\frac{\|\hat{\bI}_0\|_2}{\|\hat{\bI}_0-\mathbf{\hat{I}}\|_2}$,
where $\hat{\bI}$ is the computed hologram with axial down-sampling.
The accuracy of the BPM drops rapidly when $\Delta z$ is smaller than the particle diameter ($1\upmu\mr{m}$) and reduces slowly as $\Delta z$ further increases. 
This indicates that to accurately compute the inner-particle multiple scattering using the BPM, dense axial sampling is generally needed.
Nevertheless, even under coarse axial sampling up to $\Delta z=16\lambda$, the BPM is still more accurate than the FBM computed with the reference dense axial sampling $\Delta z=\lambda/16$.

Next, we study the BPM's accuracy for different $\Delta z$ for different particle densities $\rho$ and refractive index contrasts $\Delta n$ (Fig.~\ref{fig:3}(b-c)).
Our study shows that the shape of the curve remain the same for different scattering conditions, which suggests that it is only determined by the sampling distance $\Delta z$. 

Next, we study how the scattering strength affects the BPM's accuracy for a fixed $\Delta z=\lambda/4$.
We compute holograms at different particle densities $\rho$, refractive index contrasts $\Delta n$, and volume thicknesses $Z$. 
We then compare these holograms with the corresponding reference ($\Delta z=\lambda/16$). 
The results are summarized in Fig.~\ref{fig:3}(d-f). 
In general, the accuracy decreases as the scattering becomes stronger. 
The SNR is found to satisfy the following scaling law:
\be
\frac{\|\hat{\bI}_0\|_2}{\|\hat{\bI}_0-\hat{\bI}\|_2} 
= A(\Delta z)\Delta n^{-1}(\rho\times X \times Y\times Z)^{-0.5} 
= A(\Delta z)\Delta n^{-1}P^{-0.5},
\label{equ24}
\ee
where we define a scaling parameter $A(\Delta z)$ to describe the effects of $\Delta z$ whose values are proportional to that shown in Fig.~\ref{fig:3}(a). $X, Y, Z$ are the lateral and axial sizes of the object volume and $P$ is the total number of particles. 
Intuitively, Eq.~\eqref{equ24} shows that the SNR is approximately inversely proportional to the total scattering potential $V$ (where $V=\frac{k_0^2}{3}R^3(n^2-n_0^2)P$, $n$ is the refractive index of the particles, $R=0.5\upmu\mr{m}$ is the radius of the particles) of the particle volume.

\section{3D imaging results}
In this section, we report 3D particle imaging results in both simulation and experiments. 
\begin{figure}[ht!]
 \centering 
 \includegraphics[width=1\textwidth]{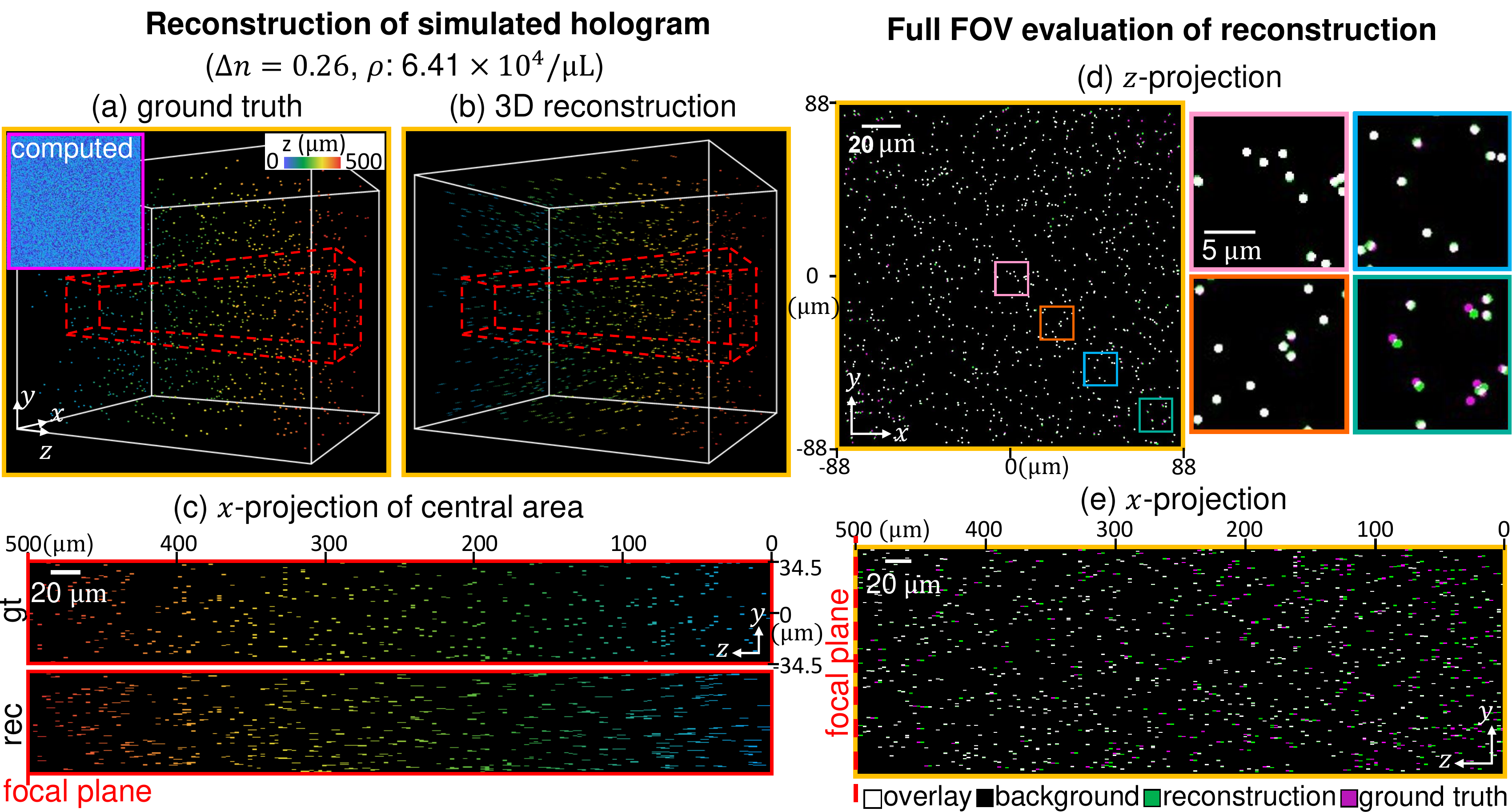}
 \caption{\label{fig:4} 3D imaging results on a simulated hologram.
 (a) The ground truth particle distribution and simulated hologram. 
 (b) 3D reconstruction of our algorithm. 
 (c) $x$-projections of the ground-truth and reconstructed particles in the central region (in red box). 
 The reconstructed particles' centroids overlaid onto the ground truth shown in (d) $z-$projection and (e) $x$-projection. For visualization, the extracted centroid of each particle is enlarged to a $1\upmu \mr{m}$ disc. Note the definition of the $z$-direction is the same as that in Fig.~\ref{fig:1}, in which the left sides of (c) and (e) lie closest to the hologram plane.
 }
\end{figure}

\subsection{Simulation results}
Given the high accuracy of the BPM model validated in Sec.~\ref{sec:3}, we next quantify the accuracy of our reconstruction algorithm in simulation. 
To implement the reconstruction algorithm, we first select a suitable $\Delta z$.
Due to memory limitations, it is not feasible to perform reconstructions using the same dense $\Delta z=\lambda/16$ as the forward simulation. 
Based on our quantitative study in Fig.~\ref{fig:3}, we select $\Delta z=6.2\lambda$ for all the reconstructions, which balances the model accuracy and computational cost.
The corresponding object volume contains around 177 million voxels.

In Fig.~\ref{fig:4}(a), we simulated a hologram from a particle field ($\rho=6.41\times10^4/\upmu\mr{L}, \Delta n =0.26$, particle diameter:$1\upmu\mr{m}$) using the densely sampled BPM model ($\Delta z = \lambda/16$).
The 3D reconstruction result is visualized in Fig.~\ref{fig:4}(b). 
For better visualization, we show the $x$-projection of the central $69\times69\times500\upmu\mr{m}^3$ region in Fig.~\ref{fig:4}(c). 
As expected, the reconstructed particles are elongated along the $z$-axis due to the missing cone problem.
To further evaluate the reconstruction, we extracted the centroids of the reconstructed particles and then overlay them onto the ground-truth.
The $z$- and $x$-projections are shown in Fig.~\ref{fig:4}(d-e).
As shown in the $z$-projection, the reconstruction errors mostly occurred in the peripheral FOV region.
Further inspecting the the $x$-projection, the localized centroids agree well with the ground truth.  
\begin{figure}[ht!]
 \centering 
 \includegraphics[width=1\textwidth]{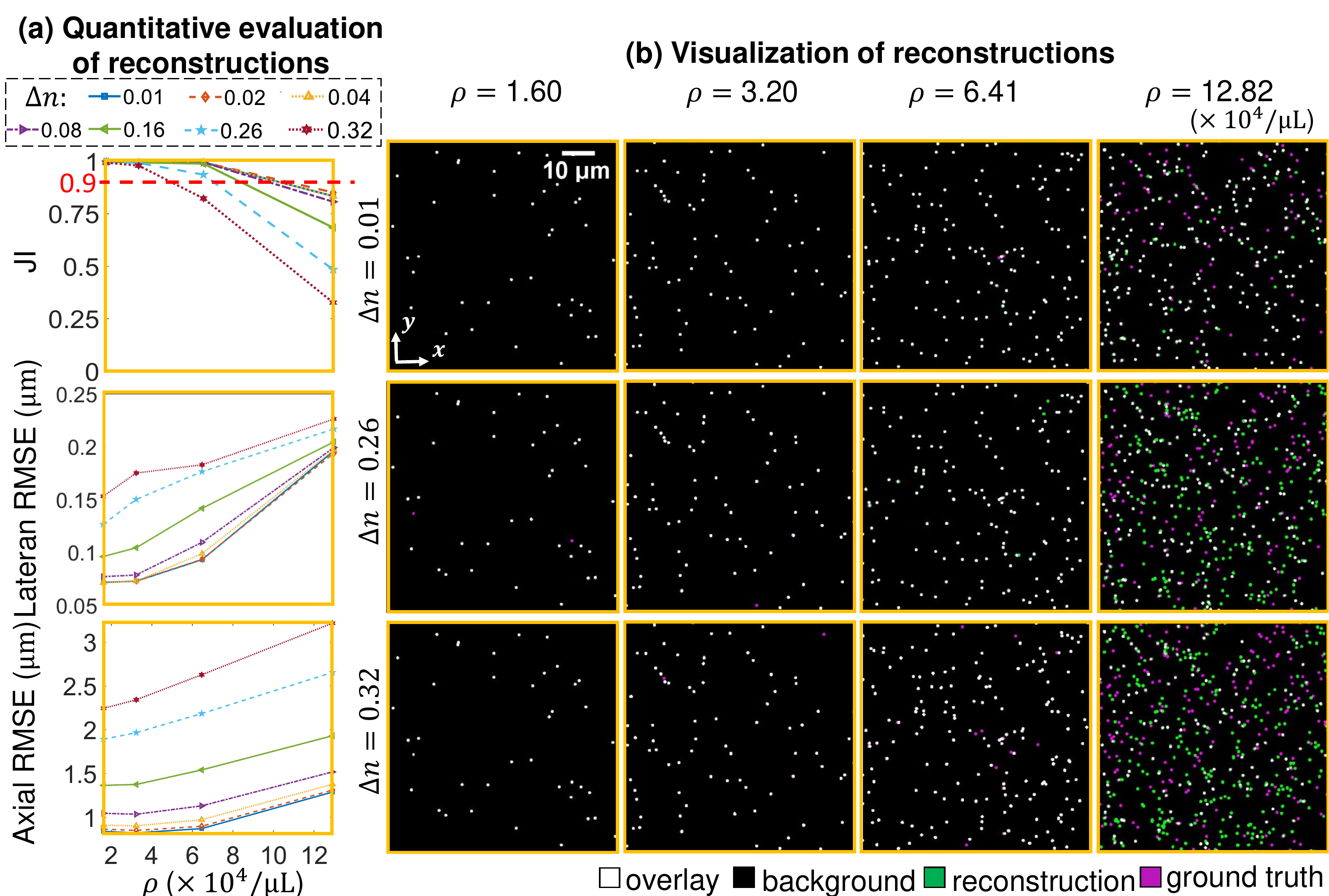}
 \caption{\label{fig:5} Quantitative evaluation of the reconstruction results at different scattering conditions. 
 (a) The JI, lateral/axial RMSE are quantified at different densities and refractive index contrasts. Each metric is averaged from 10 independent simulations. 
 (b) The $z$-projections of the reconstructed particles overlaid on the ground-truth across 4 particle densities and 3 refractive index contrasts.}
\end{figure}

To further quantify the reconstruction accuracy, we repeat the simulation at seven different refractive index contrasts and four different particle densities.
We then use Jaccard index (JI), lateral root mean square error (RMSE), and axial RMSE for quantification~\cite{sage2019super}: 
\be
\mr{JI} = \frac{\mr{TP}}{\mr{TP+FP+FN}},
\label{equ26}
\ee
\be
\mr{Lateral\, RMSE} =  \sqrt{\frac{1}{\mr{TP}}\sum_{i\in B}(\delta x^2+\delta y^2)},
\label{equ27}
\ee
\be
\mr{Axial\, RMSE} =  \sqrt{\frac{1}{\mr{TP}}\sum_{i\in B}\delta z^2},
\label{equ28}
\ee
where TP, FP, and FN denote the number of true positive, false positive, and false negative particles, respectively,
$\delta x, \delta y,$ and $\delta z$ measure the distance between the centroids of the reconstructed and the matching ground-truth particle, and $B$ is the set of all TP particles, see more details in Supplementary material.

As shown in Fig.~\ref{fig:5}(a), our method achieves JI > 0.9 for imaging particles at densities $\rho \le 6.41\times10^{4} /\upmu \mr{L}$ (1000 particles in the  $176.74\times176.74\times500 \upmu \mr{m}^3$ volume) with $\Delta n = 0.26$. 
Our method also achieves localization accuracy better than the diffraction limit.
The lateral RMSE is less than 0.25$\upmu \mr{m}$ and the axial RMSE is less than 3.5$\upmu \mr{m}$ in all cases.
Figure~\ref{fig:5}(b) shows representative $z$-projections of the centorids overlaid onto the ground-truth for the central $69\times69\times500\upmu \mr{m}^3$ region. 
These results show that our method performs well for particle densities as high as $6.41\times10^4 /\upmu \mr{L}$ and refractive index contrast as high as 0.32.
\begin{figure}[ht!]
 \centering 
 \includegraphics[width=1\textwidth]{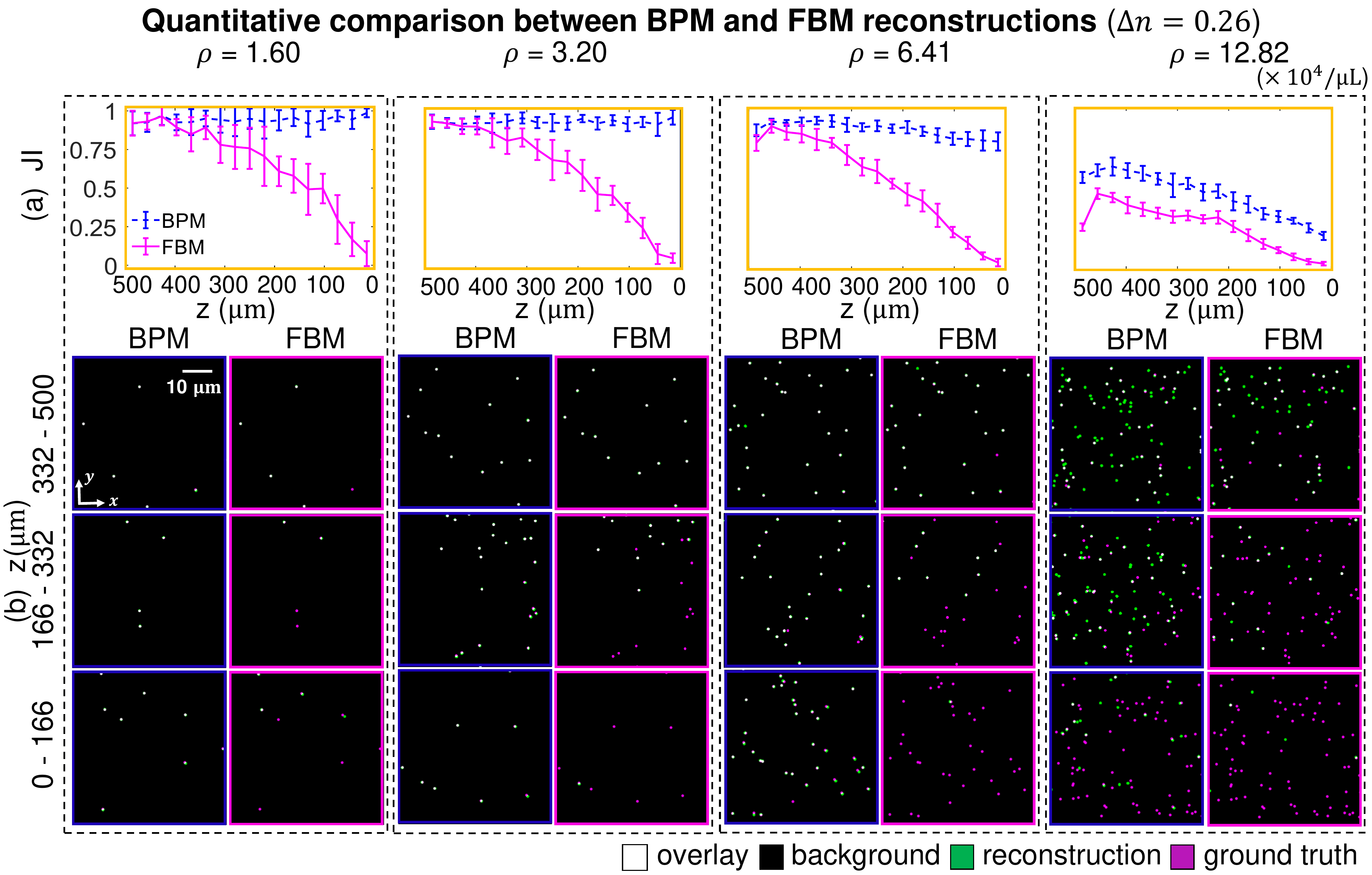}
 \caption{\label{fig:6} Quantitative comparison between BPM and FBM reconstruction algorithms. 
 (a) The depth-dependent JIs for the BPM and FBM reconstructions under four different particle densities show superior performance by the BPM method.
 (b) The $z$-projections of the BPM and FBM reconstructions overlaid on the ground truth at 3  depth regions and 4 particle densities.}
\end{figure}

Next, we quantify the improvement of our multiple-scattering BPM model as compared to the single-scattering FBM model.
To do so, we performed reconstructions using the compressive holography algorithm that uses the FBM forward model and a total variation (TV) regularization to enforce sparsity~\cite{brady2009compressive,chen2015empirical}. 
In Fig.~\ref{fig:6}, we compare results across four different particle densities with $\Delta n = 0.26$.
To quantify the depth-dependent reconstruction accuracy, we calculated JI by dividing the entire depth into 17 segments and then calculating JI for each segment.
The depth-wise JIs under different scattering densities are shown in Fig.~\ref{fig:6}(a).
The results show that the BPM-based algorithm consistently detects the particles across all the depths for particle densities $\rho \le 6.41\times 10^{4} /\upmu \mr{L}$.
In contrast, the FBM-based algorithm suffers from rapid degradation as the depth increases.
In particular, when the depth is around $500 \upmu \mr{m}$, the JI of BPM is $>34$ times higher than the FBM when $\rho \le 6.41\times 10^{4} /\upmu \mr{L}$.
Representative $z$-projections of the reconstructed centroids overlaid onto the ground truth across different depths are shown in Fig.~\ref{fig:6}(b) for the central $69\times69\times500 \upmu \mr{m}^3$ region.
These results highlight that our multiple-scattering algorithm significantly outperforms the traditional single-scattering method, in particular for reconstructing particles at deep depths. 

\subsection{Experimental results}
\begin{figure}[ht!]
 \centering 
 \includegraphics[width=1\textwidth]{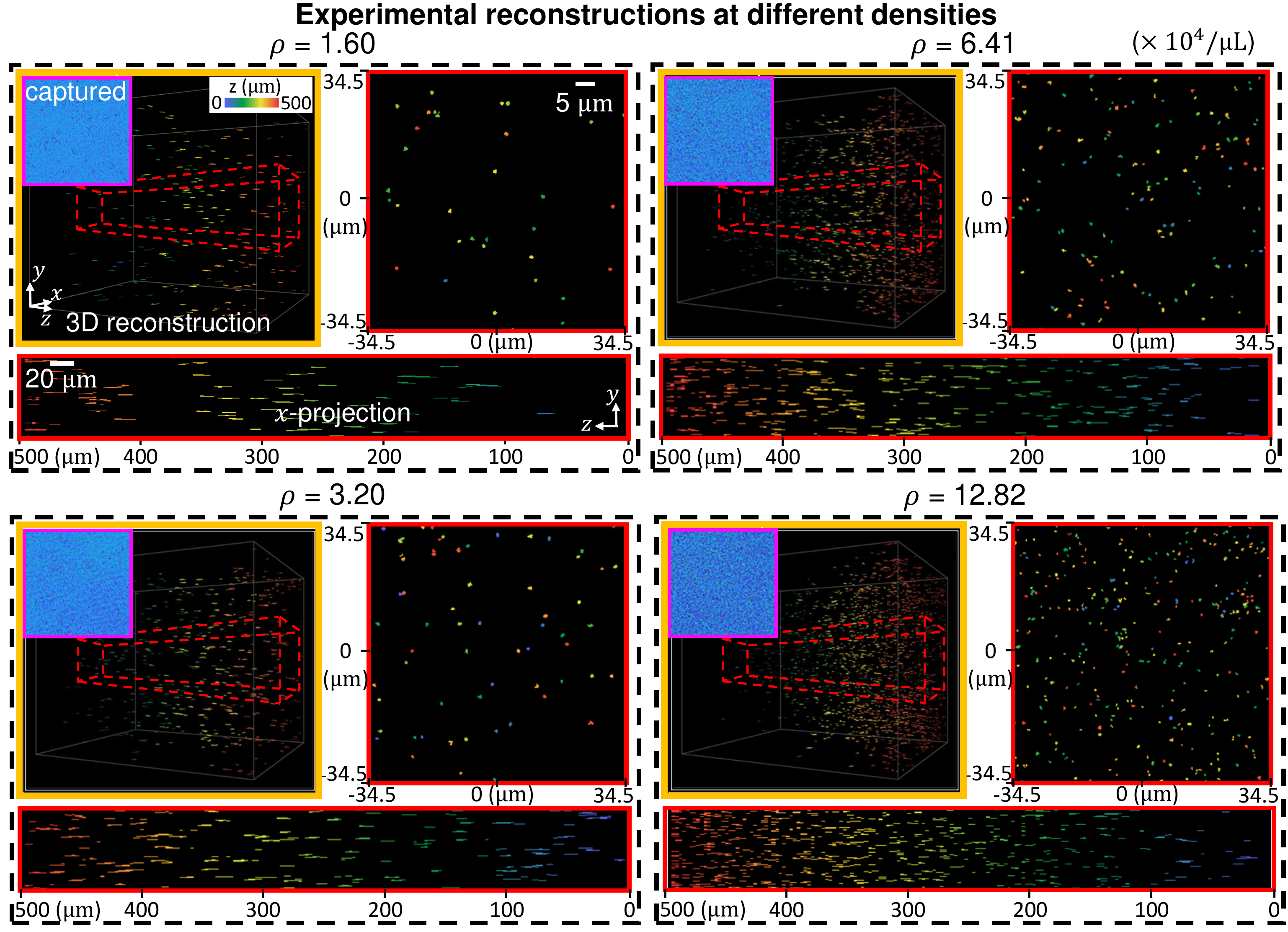}
 \caption{\label{fig:7} Experimental reconstruction results at four different particle densities. 
 The captured holograms are shown in the top-left insets in each sub-figure.
 The 3D rendering of the whole volume and orthogonal depth-color coded projections of the sub-volume outlined in red are shown for each reconstruction.
 }
\end{figure}

We demonstrate our reconstruction algorithm on experimentally captured holograms at 4 different densities. 
In Fig.~\ref{fig:7}, 3D visualizations of example reconstruction results are shown in depth-color coded 3D renderings and $x$- and $z$-projections.
The axial elongations are visible in all cases, which are resulted from the missing spatial frequency information limited by the single-view holographic measurement.
We further observe that more particles are reconstructed at the depths closer to the hologram plane. 
This is expected since, for particles closer to the hologram plane, a greater amount of scattered information are captured by the finite-sized image sensor.

\begin{figure}[ht!]
 \centering 
 \includegraphics[width=1\textwidth]{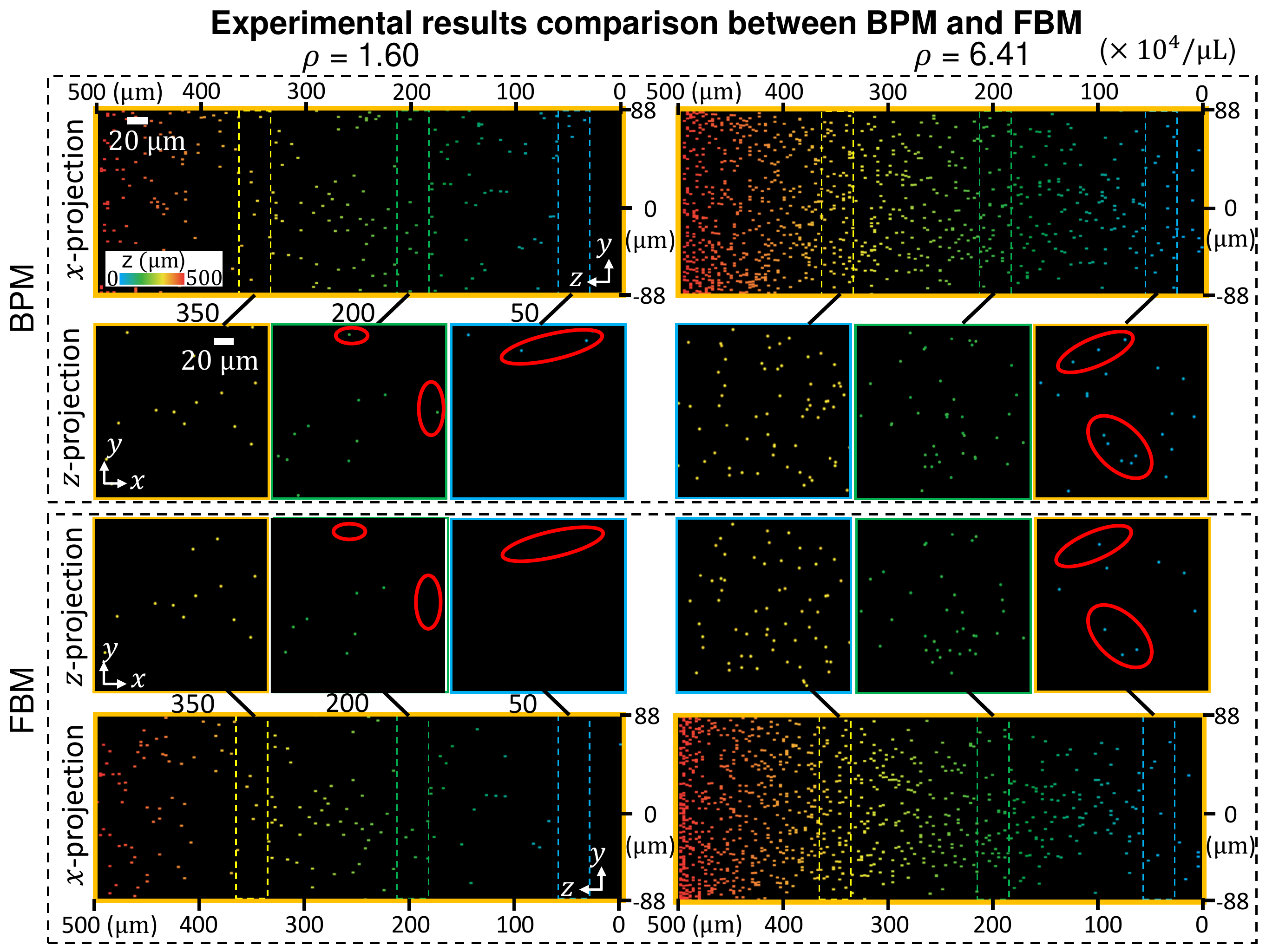}
 \caption{\label{fig:8} Experimental comparisons between the BPM and FBM reconstructions. The $x$-projection of the entire volume and multiple $z$-projections at different depth ranges of the reconstructed centroids are shown. The BPM outperforms the FBM particularly for reconstructing particles at deep depths, as highlighted by the red circles.}
\end{figure}

Finally, we compare our method with the FBM algorithm in Fig.~\ref{fig:8}.
Similar to our observations in the simulation, the BPM algorithm can better reconstruct particles at deeper depths than the FBM algorithm, as highlighted by the $x$- and $z$-projections for different depth regions.

\section{Conclusion}
We have developed a new reconstruction algorithm for large-scale holographic particle 3D imaging based on a multiple-scattering beam propagation model. 
Our forward model demonstrates superior accuracy for multiple-scattering particle fields as compared to the traditional first Born-based single scattering model.
The computational efficiency of our iterative algorithm allows reducing the computational complexity by more than 2 orders of magnitude for reconstructing volumes containing morn than 100 million voxels, as compared to the iterative Born series based method.
The accuracy of our proposed algorithm is demonstrated in both simulation and experiment.
In particular, we show that our method is particularly effective to improve the imaging performance for particles at deep depths.
Together, these advances may open up new exciting opportunities for large-scale holograph particle 3D imaging in various applications.

Though our algorithm achieved start-of-the-art performance, it is still limited by the severe missing cone problem, which elongates the reconstructed particles. 
A promising future direction is to combine our multiple-scattering model and advanced deep-learning priors to further improve the imaging performance~\cite{wu2020simba, matlock2021physical}. 

\begin{backmatter}
\bmsection{Funding}
National Science Foundation: 1813848 and 1846784.

\bmsection{Acknowledgments}
We thank Boston University
Shared Computing Cluster for the computing resources.

\bmsection{Disclosures}
The authors declare that there are no conflicts of interest related to this article.

\bmsection{Data availability}
Data underlying the results presented in this paper are available in Ref~\cite{codeexample}.

\end{backmatter}

See Supplement 1 for supporting content.
\bibliography{3Dlocalozation.bib}
\end{document}